\documentclass[prl,superscriptaddress,twocolumn]{revtex4-1}
\usepackage{times,amsmath}
\usepackage{epsfig}
\usepackage{graphicx}
\usepackage{dcolumn}
\usepackage{bm}
\usepackage{tabularx}
\usepackage{hyperref}
\hypersetup{citecolor=black, filecolor= black, linkcolor= black, urlcolor= black}
\begin{document}

\preprint{version 0}

\title{Systematic search for low-enthalpy $sp^3$ carbon allotropes using evolutionary metadynamics}

\author{Qiang Zhu}
\email{qiang.zhu@stonybrook.edu}
\affiliation{Department of Geosciences, Department of Physics and Astronomy, Stony Brook University, Stony Brook, New York 11794, USA}
\author{Qingfeng Zeng}
\affiliation{Department of Geosciences, Department of Physics and Astronomy, Stony Brook University, Stony Brook, New York 11794, USA}
\affiliation{National Key Laboratory of Thermostructure Composite Materials, Northwestern Polytechnical University, Xi'an 710072, PR China}
\author{Artem R. Oganov}
\affiliation{Department of Geosciences, Department of Physics and Astronomy, Stony Brook University, Stony Brook, New York 11794, USA}
\affiliation{Geology Department, Moscow State University, Moscow 119992, Russia}

\begin{abstract}
We present a systematic search for low-energy metastable superhard carbon allotropes by using the recently developed evolutionary metadynamics technique. It is known that cold compression of graphite produces an allotrope at 15-20 GPa. Here we look for all low-enthalpy structures accessible from graphite. Starting from 2H- or 3R-graphite and applying a pressure of 20 GPa, a large variety of intermediate $sp^3$ carbon allotropes were observed in evolutionary metadynamics simulation. Our calculation not only found all the previous proposed candidates for `superhard graphite', but also predicted two allotropes (\emph{X}-carbon and \emph{Y}-carbon) showing unusual types of 5+7 and 4+8 topologies. These superhard carbon allotropes can be classified into five families based on 6 (diamond/lonsdaleite), 5+7 (\emph{M}- and \emph{W}-carbon), 5+7 (\emph{X}-carbon), 4+8 (bct-C$_4$), and 4+8 (\emph{Y}-carbon) topologies. This study shows evolutionary metadynamics is a powerful approach both to find the global minima and systematically search for low-energy metastable phases reachable from given starting materials.
\end{abstract}

\maketitle

\section{Introduction}
Carbon can adopt a wide range of structures, from superhard/superdense insulating (diamond, lonsdaleite, hypothetical phases \emph{hP}3, \emph{tI}12 and \emph{tP}12 \cite{QZhu-PRB-2011}) to ultrasoft semi-metallic (graphite, fullerenes) and even superconducting (doped diamond \cite{Ekimov-Nature-2004} and alkali-doped fullerenes \cite{Tanigaki-Nature-1991}). The quest for carbon materials with desired properties is of great interest in both fundamental science and advanced technology. One important direction in carbon research is the discovery of carbon allotropes with advanced mechanical and electronic properties. 

It is well known that graphite transforms to the thermodynamically stable diamond at high pressures ($>$ 12 GPa) and high temperatures (1900-2500 K) \cite{Irifune-Nature-2003}. On the contrary, several experiments reported that cold compression of graphite produces a metastable superhard and transparent phase, clearly different from diamond or lonsdaleite, but the exact crystal structure could not be determined \cite{Aust-Science-1963, Hanfland-PRB-1989, Zhao-PRB-1989, Yagi-Science-1991, Mao-Science-2003}. The difficulty to experimentally resolve the crystal structure has stimulated theoretical efforts \cite{Oganov-JCP-2006, Li-PRL-2009, Wang-PRL-2011, Daniele-PRB-2011, Zhao-PRL-2011, Maximilian-PRL-2012, Chen-PRL-2012, Baughman-1997, Zhou-PRB-2010, Umemoto-PRL-2010, Greshnyakov-2011}. Several structural models were found using different techniques. The physical properties of these models (\emph{M}-carbon \cite{Oganov-JCP-2006, Li-PRL-2009}, \emph{W}-carbon \cite{Wang-PRL-2011}, \emph{oC}16 (also called \emph{Z}-carbon) \cite{Daniele-PRB-2011, Zhao-PRL-2011, Maximilian-PRL-2012}, \emph{R/P} carbon \cite{Chen-PRL-2012}, bct-C$_4$ \cite{Baughman-1997, Zhou-PRB-2010, Umemoto-PRL-2010}) have been intensely studied. Simulated x-ray diffraction patterns and band gaps of these models are mostly in good agreement with experimental data, making it even harder to decide which one is the metastable product observed in experiments. On the other hand, it is not guaranteed that there is not even a better solution for this experimental puzzle. Furthermore, it is likely that different metastable phases will be obtained by room-temperature compression of different polytypes of graphite, or under various non-hydrostatic conditions. This motivates us to do a systematic search for low-energy metastable carbon allotropes.
 
So far, there are several methods to find the ground state structures of unknown materials. However, none of them are designed to search for metastable states. Our recently proposed evolutionary metadynamics method \cite{QZhu-metadynamics} can focus on that task. Starting from a reasonable initial crystal structure, with this technique one can produce efficiently both the ground state and metastable states accessible from that initial structure. In this paper, we applied this technique to systematically search for metastable carbon allotropes accessiable from graphite. Starting the calculation at 20 GPa from two polytypes of graphite (2H and 3R), we easily found the diamond structure (ground state) and a number of low-energy metastable structures with \emph{sp}$^3$ hybridization which could possibly explain `superhard graphite'. 

\section{Methodology}

The idea of metadynamics is to introduce a history-dependent potential term, which fills the minima in the free energy surface so that the system could cross the energy barriers and undergo phase transitions \cite{Laio-PNAS-2002}. This technique is usually applied as an extension of molecular dynamics (MD) simulation technique \cite{Martonak-PRL-2003}. Since it relies on MD simulation to equilibrate the system, it often leads to trapping in metastable states and amorphization rather than a transition to a stable crystal structure. We recently proposed a hybrid method, basically a metadynamics-like method driven not by local MD sampling, but by efficient global optimization moves \cite{Oganov-JCP-2006, Lyakhov-CPC-2010}. 

In this approach \cite{QZhu-metadynamics}, we start from one known initial structure at a given external pressure \emph{P}. Following Martonak \emph{et al.}, we used the cell vectors matrix $h_{ij}$ (also representable as a 6-dimensional vector \emph{h}) as a collective variable to distinguish the change of state of the system \cite{Martonak-PRL-2003}. For a given system with volume \emph{V} under external pressure \emph{P}, the derivative of the free energy \emph{G} with respect to \emph{h} is
\begin{equation} \label{eq:Gibbs1}
-\frac{\partial G}{\partial h_{ij}} = V[h^{-1}(P-p)]_{ji}
\end{equation}

At each generation (or metastep), many structures are produced and relaxed at fixed \emph{h}, and we select the lowest energy structure and compute it internal tensor \emph{p}. The technique used here generates many structures at each metastep, while in traditional metadynamics \cite{Martonak-PRL-2003}, only one structure is produced at each metastep. The cell shape is then updated with a stepping parameter $\delta$\emph{h}
\begin{equation} \label{eq:tensor}
h_{im}(t+1) = h_{im}(t)+\frac{\delta h}{|f|V^{1/3}}S_{ijkl}f_{kl}h_{jm}(t)
\end{equation}

where $S$ is the elastic compliance tensor corresponding to an elastically isotropic medium with Poisson ratio 0.26, which corresponds to the border between brittle and ductile materials \cite{Pugh-1954} and is a good average value to describe both metals and insulators. The driving force $f = -\frac{\partial G}{\partial h}$ in Eq. (\ref{eq:tensor}) comes from a history-dependent Gibbs potential $G^t$ where a Gaussian has been added to $G(h)$ at every point $h(t')$ already visited in order to discourage it from being visited again,
\begin{equation} \label{Gibbs2}
G(t) = G^h + \sum We^{-\frac{|h-h(t')|^2}{2\delta{h^2}}}
\end{equation}

where \emph{W} is the Gaussian height. Then we compute the vibrational modes for the selected structure according to the dynamical matrix constructed from bond hardness coefficients,
\begin{equation} \label{eq:dmatrix}
\begin{aligned}
D_{\alpha\beta}(a,b) = &\sum_m \Big(\frac {\partial ^2}{{\partial \alpha_a^0}{\partial \beta_b^m}} \frac{1}{2} \sum_{i,j,l,n} H_{i,j}^{l,n}(r_{i,j}^{l,n}-{r_{0}}_{i,j}^{l,n})^2\Big)
\end{aligned}
\end{equation}

Here coefficients $\alpha$, $\beta$ denote coordinates $(x,y,z)$; coefficients $a, b, i, j$ describe the atom in the unit cell; coefficients $l, m, n$ describe the unit cell number; $r_{i,j}^{l,n}$ is the distance between atom $i$ in the unit cell $l$ and atom $j$ in the unit cell $n$, while ${r_{0}}_{i,j}^{l,n}$ is corresponding bond distance, and $H_{i,j}^{l,n}$ is bond hardness coefficient computed from bond distances, covalent radii and electronegativities of the atoms \cite{Lyakhov-CPC-2010}. Note that the dynamical matrix corresponds to zero wavevector (extension to non-zero wavevectors is straightforward) and unit masses.

The simulated vibrational modes are used to produce the next generation (typically 20-40 softmutated structures). To perform softmutation \cite{Lyakhov-CPC-2010, QZhu-metadynamics}, we move the atoms along the eigenvector of the softest calculated mode. One structure can be softmutated many times using different non-degenerate modes and displacements. The magnitude of the displacement ($d_{\rm max}$) along the mode eigenvector is an input parameter: With relatively small $d_{\rm max}$ and displacements represented by a random linear mixture of all mode eigenvectors, the method becomes similar to MD-metadynamics in crossing energy barriers and equilibrating the system. With large $d_{\rm max}$ along the softest mode eigenvectors, we obtain the softmutation operator \cite{Lyakhov-CPC-2010}, capable of efficiently finding the global energy minimum. 

The next generation of softmutated structures are produced and relaxed in the updated cell. Repeated for a number of generations, this computational scheme leads to a series of structural transitions and is stopped when the maximum number of generations is reached.

In this work, structure relaxations were done using density functional theory (DFT) within the generalized gradient approximation (GGA) \cite{GGA-1996} in the framework of the all-electron projector augmented wave (PAW) \cite{PAW-1994} method as implemented in the VASP \cite{VASP-1996, VASP-PAW} code. We used a plane wave kinetic energy cutoff of 550 eV for the plane-wave basis set and a Brillouin zone sampling resolution of 2$\pi$ $\times$ 0.08 \AA$^{-1}$, which showed excellent convergence of the energy differences, stress tensors and structural parameters.

\section{Results and discussions}
In a compression experiment at low temperatures (low enough to preclude transition to the stable state), the product depends on the nature of the starting materials, and on the energy landscape (in particular, energy barriers). To fully investigate the possible candidate materials, we performed simulations starting from two different graphite polytypes (graphite-2H and 3R), which differ in the stacking of graphene layers.
\subsection{Starting from graphite-2H}
We did a preliminary test at 20 GPa starting from the graphite-2H structure, and successfully found diamond as the ground state, and \emph{M}-carbon and bct-C$_4$ as metastable states \cite{QZhu-metadynamics}. In the calculation we set $d_{\rm max}$=2.5 \AA, $W$=4000 kbar$\cdot$\AA$^3$ and $\delta{h}$=0.6 \AA. 
\begin{figure}
\epsfig{file=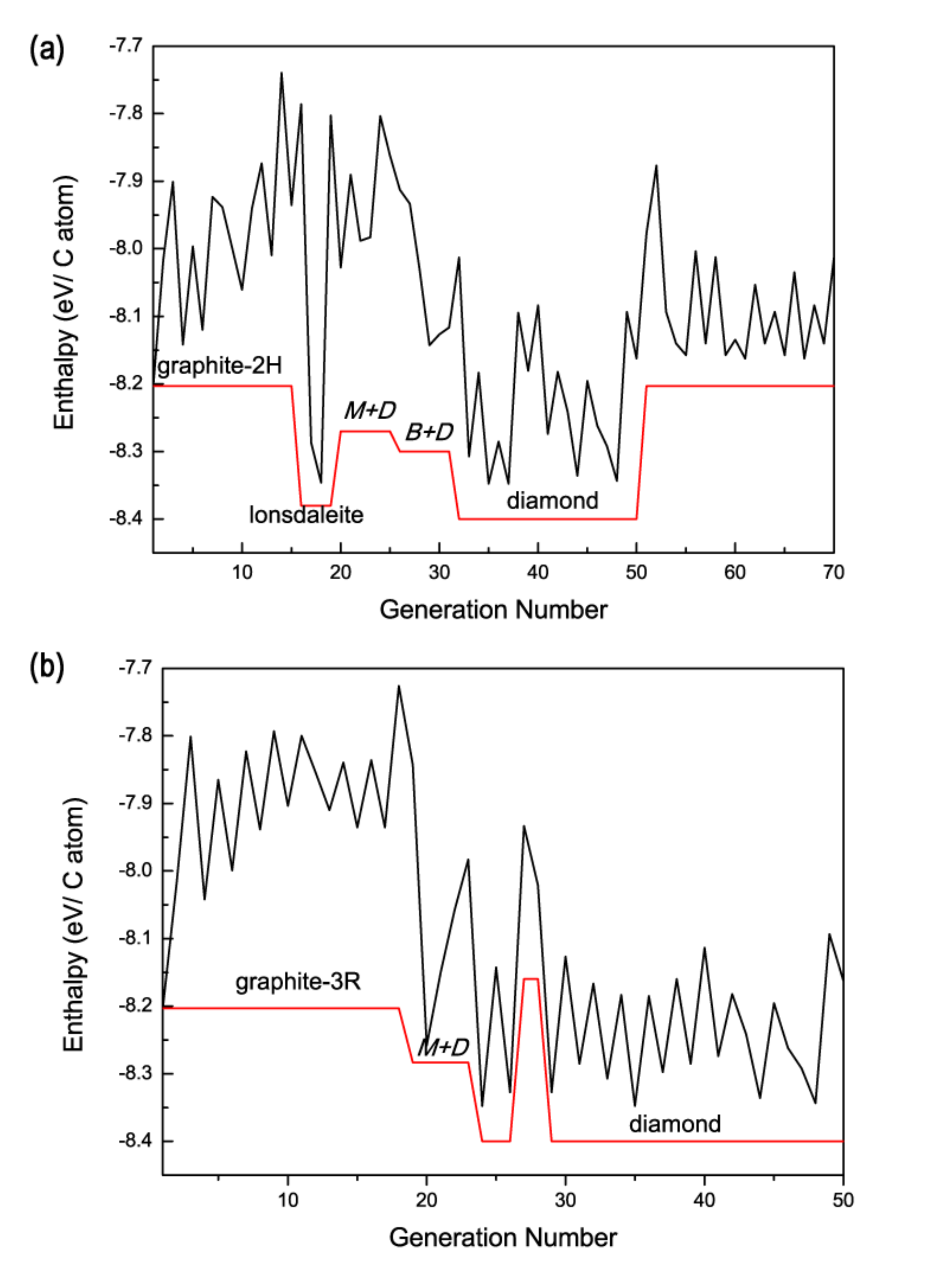, width=0.42\textwidth}
\caption{\label{C} (Color online) (a) Enthalpy evolution during the compression of graphite-2H at 20 GPa; (b) Enthalpy evolution during the compression of graphite-3R at 20 GPa (black line: enthalpies for best structures with constant cell matrix; red line: enthalpies for best structures after full relaxation).}
\end{figure}

\begin{figure}
\epsfig{file=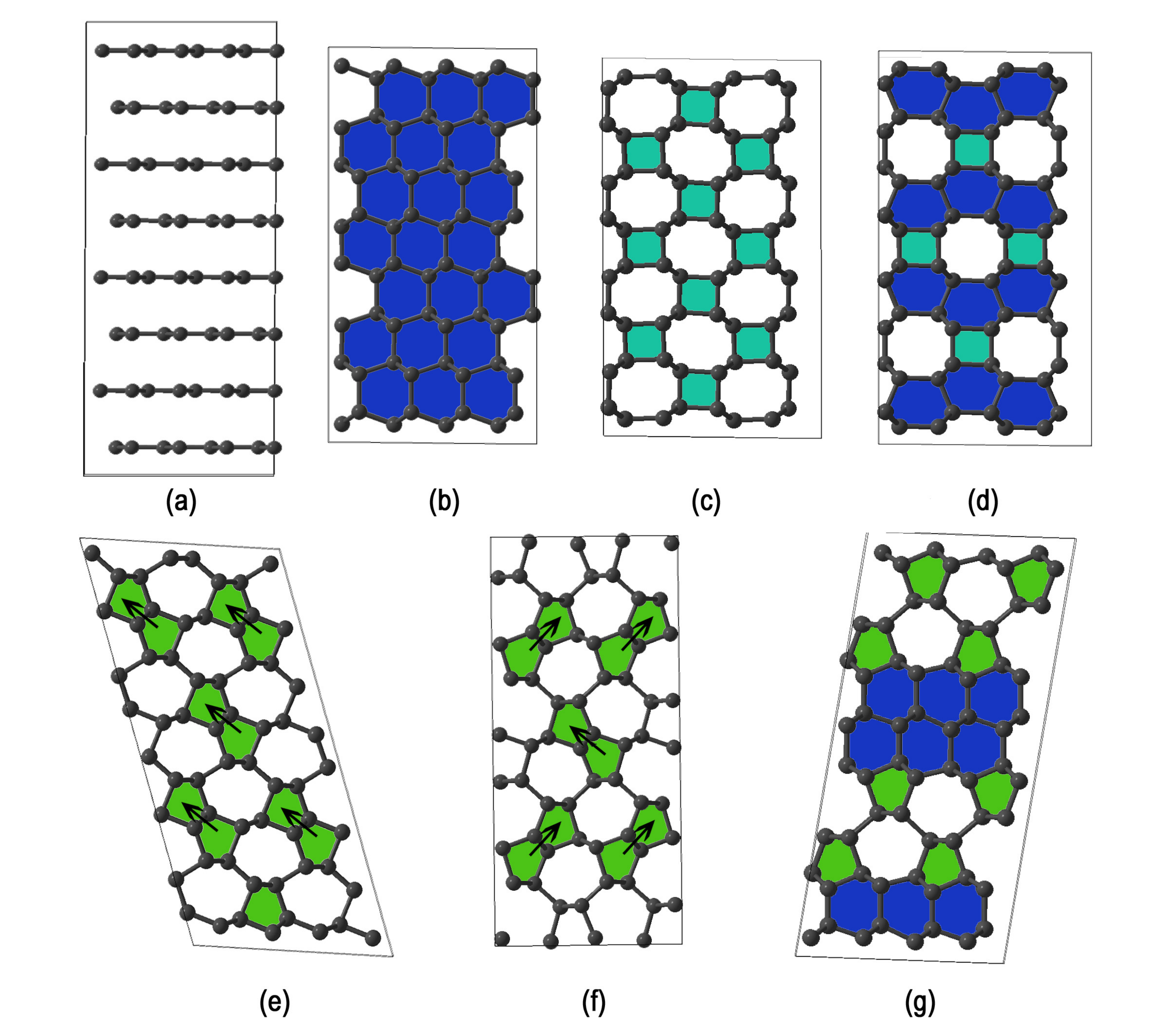, width=0.42\textwidth}
\caption{\label{2H} (Color online) Structures observed during compression of graphite-2H at 20 GPa. (a) graphite-2H (2 $\times$ 2 $\times$ 2 supercell of the calculation model); (b) lonsdaleite; (c) bct-C$_4$ with 4+8 membered rings; (d) \emph{Z}-carbon (belonging to \emph{B+D} type); (e) \emph{M}-carbon with 5+7 membered rings; (f) \emph{W}-carbon with 5+7 membered rings; (g) \emph{M+D} type carbon. Polygons are highlighted by different colors (quadrangle: turquoise; pentagon: green; hexagon: blue).}
\end{figure}

Fig. \ref{C} shows the enthalpy evolution. Graphene layers (Fig. \ref{2H}a), persisted until the 15th generation. Then, upon sufficient cell deformation, the layers began to buckle, and the planar structure transformed into 3D-networks of \emph{sp}$^3$-hybridized carbon atoms. Lonsdaleite with 6-membered rings (Fig. \ref{2H}b) appeared as the best structure in the 16th generation. We observed in the same generation the bct-C$_4$ structure with 4+8 membered rings (Fig. \ref{2H}c), and \emph{M/W}-carbon structures containing 5+7 membered rings (Fig. \ref{2H}e,f), appeared shortly after. Lonsdaleite survived for a few generations until it tranformed into a hybride structure made of alternating layers of \emph{M}-carbon and diamond (Fig. \ref{2H}g, we refer to it as \emph{M+D} type), followed by the transition to another hybride structure made of bct-C$_4$ and diamond (Fig. \ref{2H}d, similarly, we refer to it as \emph{B+D} type). Diamond was dominant in the following generations. At the 51st generation, the system reverted to graphite.

The power of the evolutionary metadynamics method lies in that it is highly suitable for harvesting low-energy metastable structures in addition to the ground state. Those previously proposed candidate structures for the product of cold compression of graphite, bct-C$_4$, \emph{M}, and \emph{W}-carbon are all easily recovered in a single simulation. More interestingly, we also observed many low-energy structures based on 5+7 or 4+8 topology. The \emph{B+D} type structure (Fig. \ref{2H}d) observed in the simulation is actually the \emph{oC}16 structure (sometimes called \emph{Z} carbon), recently suggested as a candidate for superhard graphite. Since \emph{oC}16 inherits layers of bct-C$_4$ and diamond, there is no surprise that its thermodynamic properties are intermediate between these two structures. The 5+7 class of structures shows a larger diversity. In some of these structures, 5-membered rings form pairs, while in others these they are single. The difference of the 5-membered ring pairs' orientation leads to two allotropes: \emph{M}-carbon (Fig. \ref{2H}e) and \emph{W}-carbon (Fig. \ref{2H}f). Some structures can be thought of as combinations of layers of the \emph{M}-carbon and diamond structures (\emph{M+D} carbon, as shown in Fig. \ref{2H}g)

\subsection{Starting from graphite-3R}
\begin{figure}
\epsfig{file=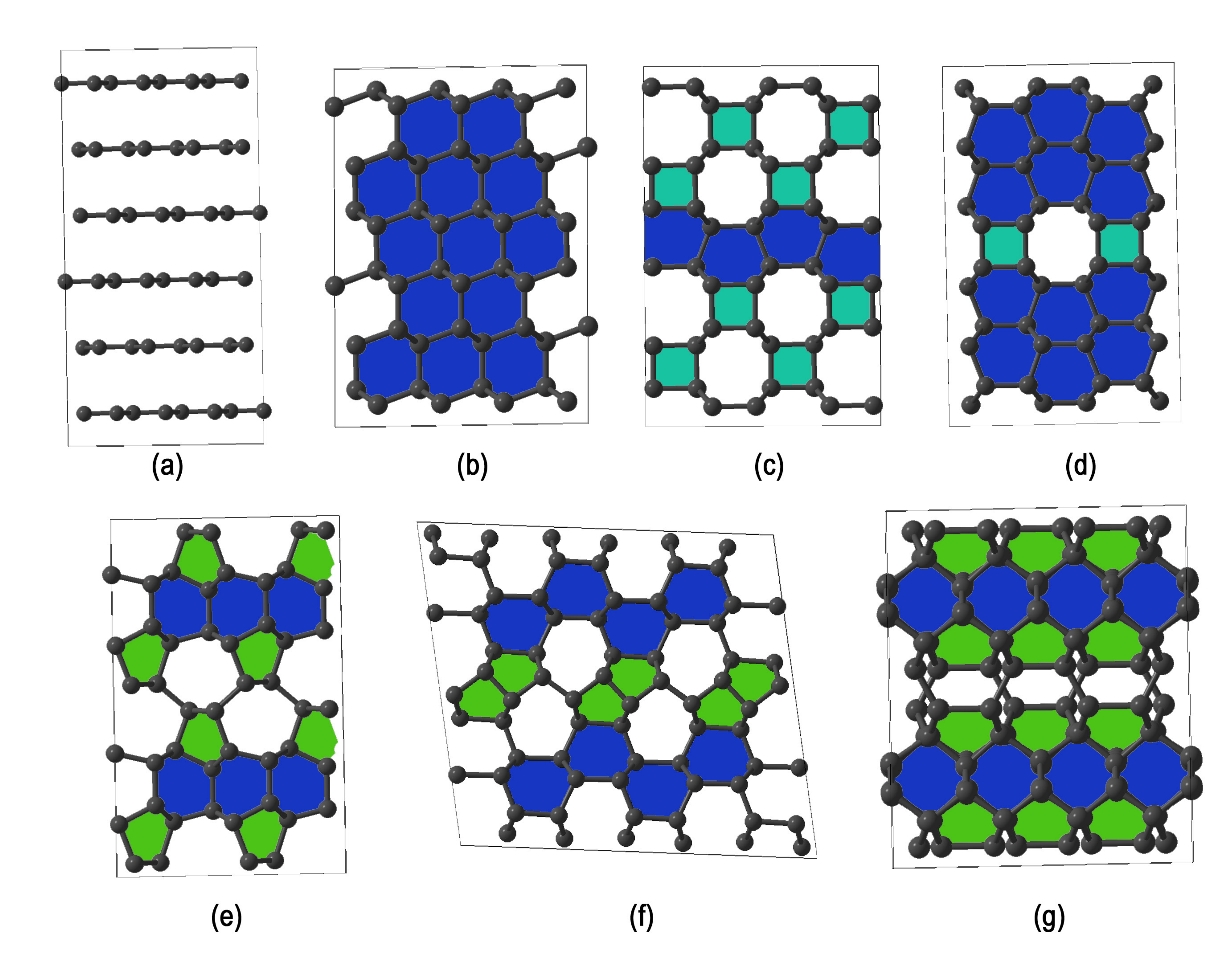, width=0.42\textwidth}
\caption{\label{3R}(Color online) Structures observed during compression of graphite-3R at 20 GPa. (a) graphite-3R (2 $\times$ 2 $\times$ 2 supercell of the calculation model); (b) diamond; (c,d) \emph{B+D} type structures; (e,f) \emph{M+D} type structures; (g) \emph{X+D} type structures. Polygons are highlighted by different colors (squares: turquoise; pentagons: green; hexagons: blue).}
\end{figure}

\begin{figure}
\epsfig{file=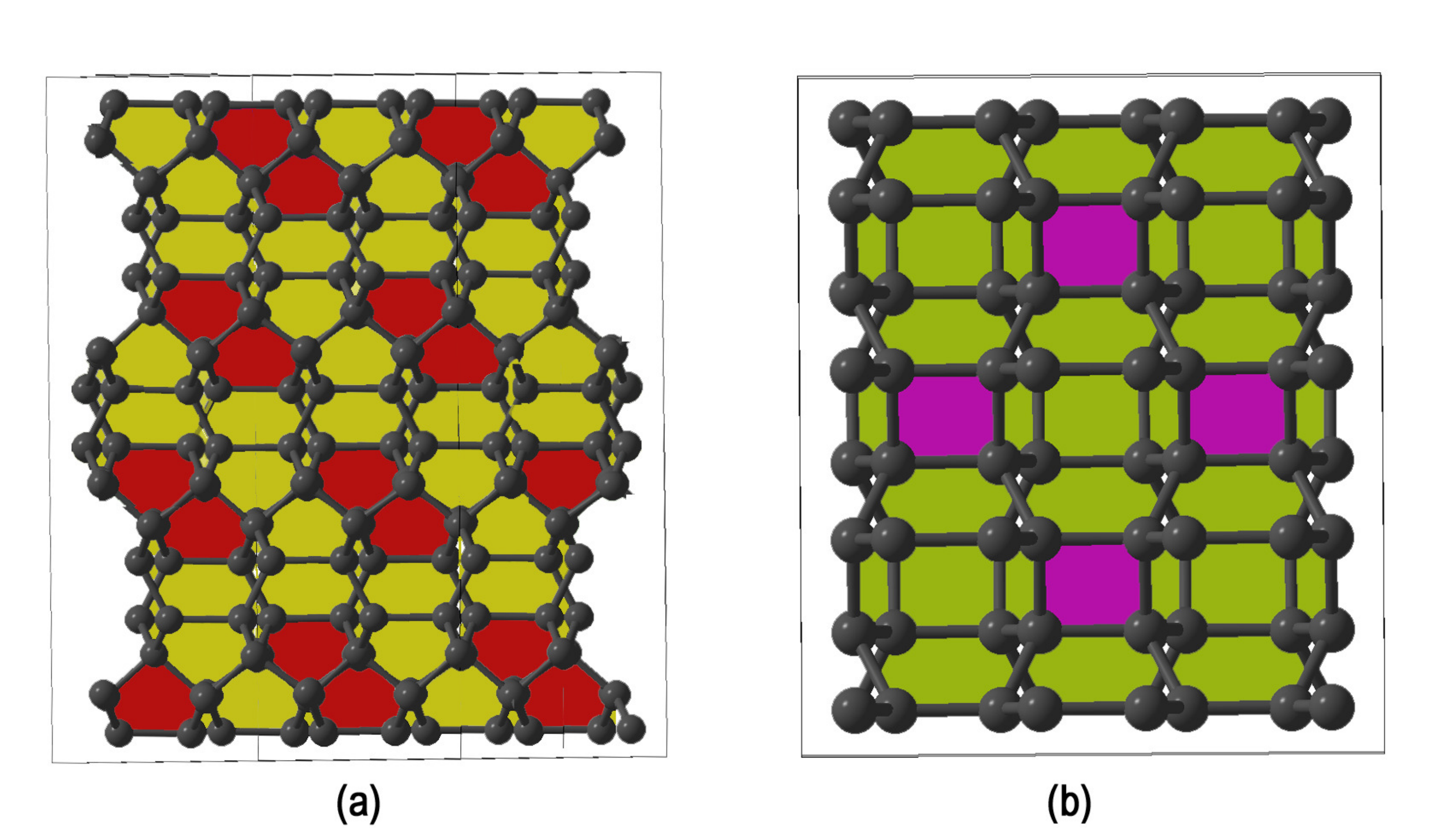, width=0.4\textwidth}
\caption{\label{5+7}(Color online) (a) New allotropes with 5+7 topology, \emph{X}-carbon, space group \emph{C}2/\emph{c}, a=5.559 \AA, b=7.960 \AA, c=4.752 \AA, $\beta$=114.65$^\circ$. This structure has five non-equivalent Wyckoff positions: C1(0.250, 0.083, 0.949), C2(0.489, 0.809, 0.982), C3(0.000, 0.200, 0.250), C4(0.247, 0.913, 0.801), C5(0.000, 0.816, 0.250). (b) New allotrope with 4+8 topology, \emph{Y}-carbon, space group \emph{Cmca}, a=4.364 \AA, b=5.057 \AA, c=4.374 \AA. This structure has one Wyckoff position: C(0.681, 0.635, 0.410). Polygons are highlighted in different colors to show the 5+7 and 4+8 topology.}
\end{figure}

\begin{figure}
\epsfig{file=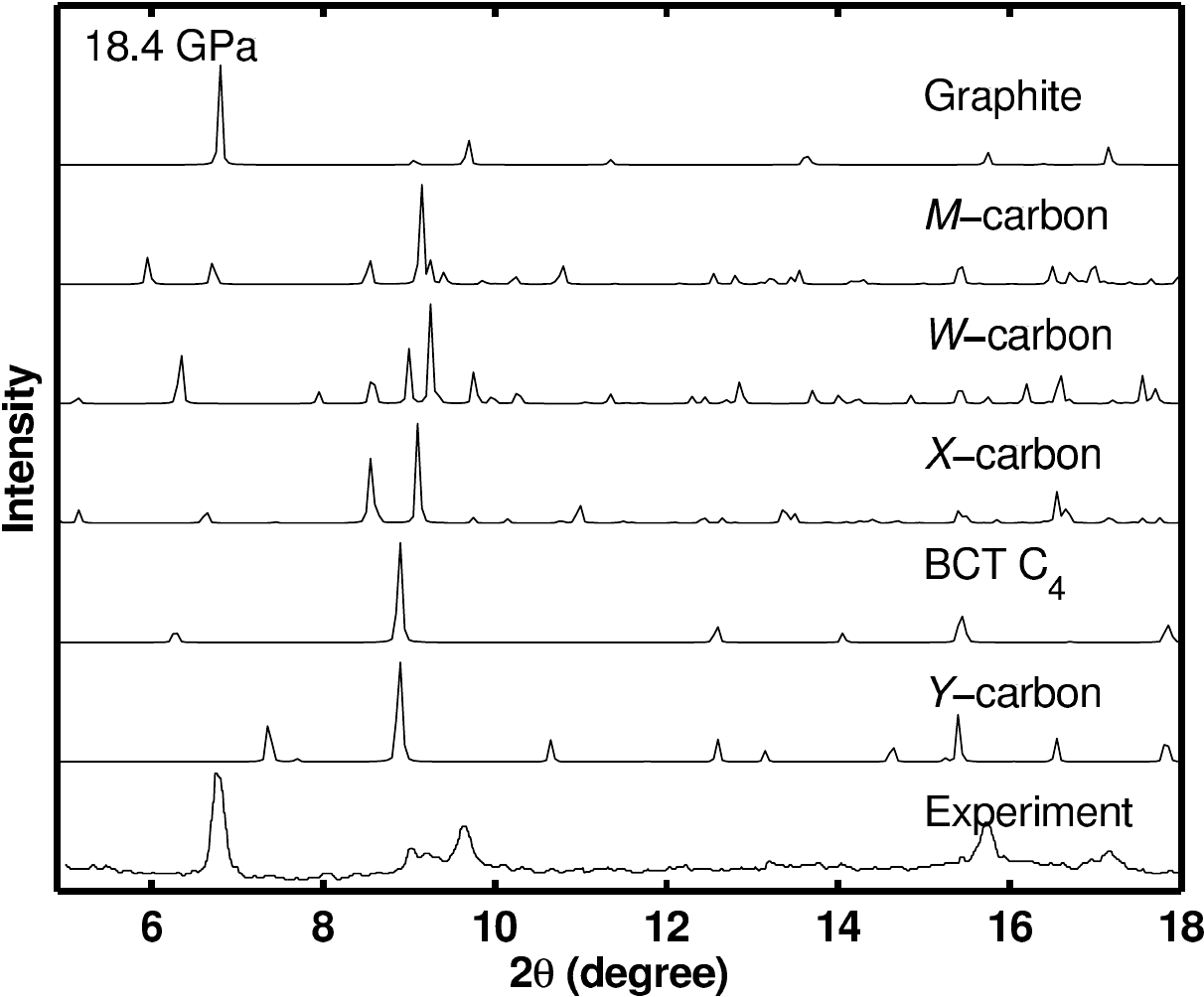, width=0.35\textwidth}
\caption{\label{xrd} Comparison of simulated x-ray diffraction patterns of proposed models and graphite with experiment \cite{Mao-Science-2003}. The patterns were simulated using Accelrys Materials Studio 4.2 software with the X-ray wavelength of 0.3329 \AA.}
\end{figure}

Starting the calculation at 20 GPa from another polytype, graphite-3R, which contains three layers per lattice period, we again easily found the diamond structure and a number of low-energy metastable structures with \emph{sp$^3$} hybridization. Fig. \ref{3R} shows the results. Since the initial model has three graphene layers, it could form a large variety of \emph{M+D} and \emph{B+D} structures based on 4+6+8 or 5+6+7 topologies. For instance, we observed a \emph{B+D} structure containing 2$\times$(4+8) layers and 1$\times$6 layer (Fig. \ref{3R}c), or 1$\times$4+8 layers and 1$\times$6 layer (Fig. \ref{3R}d); and \emph{M+D} structure containing 2$\times$(5+7) layers and 1$\times$6 layer (Fig. \ref{3R}e,f). Most strikingly, we also observed another structure with a 5+7 topology, which is in Fig. \ref{3R}g. The projections of pentagons and heptagons along the \emph{c} axis could not be separated as in \emph{M}-carbon, but overlap each other. We extracted the 5+7 part from the complex structure, and obtained a different configuration with pure 5+7 topology. This crystal structure (which we call \emph{X}-carbon, and the hybrid structure from \emph{X}-carbon and diamond is referred to as \emph{X+D} type) is shown in Fig. \ref{5+7}. It is a monoclinic structure with \emph{C}2/\emph{c} symmetry, and contains 32 atoms in the conventional cell. We also found an unexpected 4+8 topology in an allotrope that we call Y-carbon with unique 4+8 ring topology from another separate metadyanmics run. It is an orthorhombic structure with \emph{Cmca} symmetry, containing 16 atoms in the conventional cell. The simulated X-ray diffraction patterns of all these structures are in good agreement with experimental data (as shown in Fig. \ref{xrd}), suggesting that both \emph{X,Y} carbon could explain the experiments on cold compressed graphite. Although all these structures show a satisfactory agreement with experimental X-ray data, our recent transition path sampling calculations \cite{SEB-2012} suggest \emph{M}-carbon to be kinetically the likeliest product of cold compression of graphite-2H. Using other polytypes of graphite, or different conditions (non-hydrostatic or dynamical compression), one might produce alternative allotropes found here. Synthesis of these allotropes would be desirable in view of their physical properties.

\subsection{Properties}
\begin{figure}
\epsfig{file=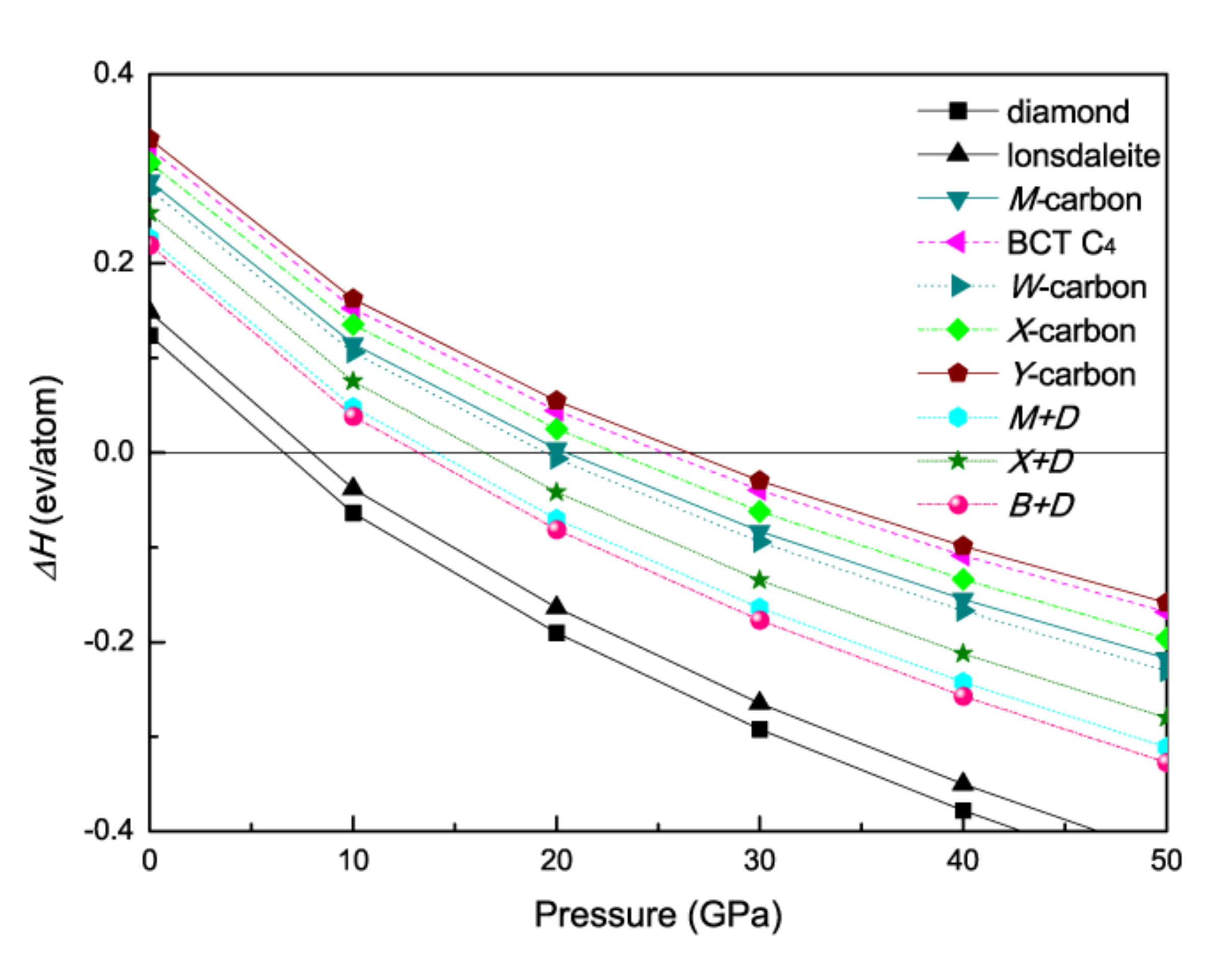, width=0.35\textwidth}
\caption{\label{E-P}(Color online) Enthalpies of various carbon structures relative to graphite-2H.}
\end{figure}

From evolutionary metadynamics simulations, five families of $sp^3$-hybridized structures made by stacking corrugated graphene layers and having competitive enthalpies were discovered: 6 (diamond and lonsdaleite), two classes of 5+7 topologies (one - \emph{M/W}-carbon; the other - \emph{X}-carbon), and two classes of 4+8 (one - bct-C$_4$; the other - \emph{Y}-carbon). The enthalpies of different carbon phases as a function of pressure are presented in Fig. \ref{E-P}. Apart from the prototypes, we also included hybrid structures (lowest enthalpy \emph{B+D}, \emph{M+D} and \emph{X+D} carbon, see crystallographic data in Supplementary Materials). At elevated pressures, all these allotropes become more stable than graphite. For the prototypes, \emph{M/W}-carbon is energetically more favorable than bct- and \emph{X}-carbon. Lower enthalpies are obtained by combining layers of these structures with layers of diamond. For the models under consideration (up to 4 graphene layers), \emph{B+D} (1$\times$(4+8) + 2$\times$ 6 layers, Fig. \ref{3R}d) tends to have the lowest enthalpy, while \emph{M+D} (2$\times$(5+7) + 2$\times$6 layers, Fig. \ref{2H}h) is quite competitive (only 8 meV/atom higher than \emph{B+D}). \emph{X+D} (Fig. \ref{3R}h) is 50 meV/atom higher than \emph{B+D}, indicating that \emph{X}-carbon has poorest possibility to interface with diamond.

\begin{figure}
\epsfig{file=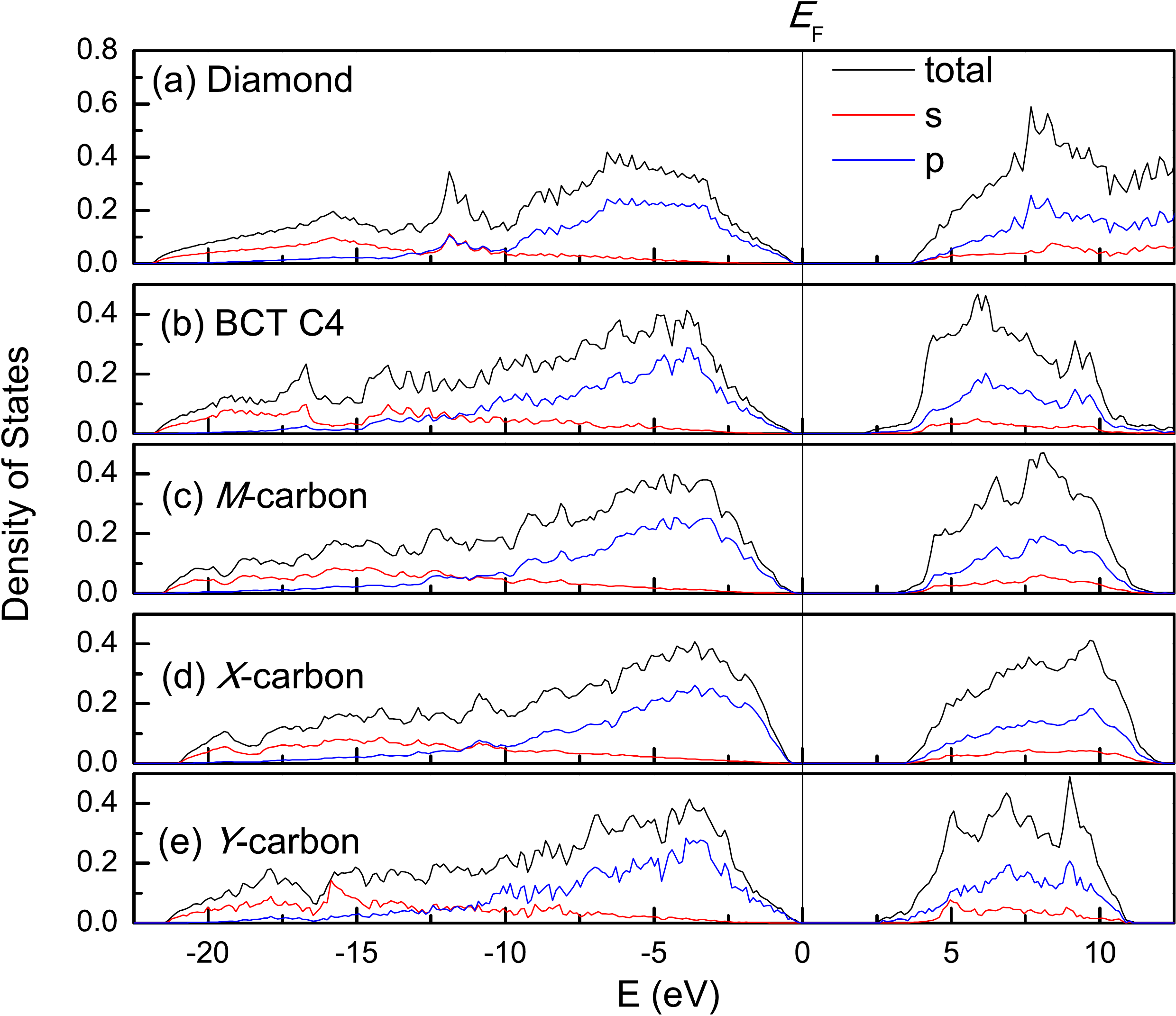, width=0.35\textwidth}
\caption{\label{DOS}(Color online) Total and partial densities of states of different carbon allotropes.}
\end{figure}

We also computed the mechanical and optical properties (see Supplementary Materials). Similar to previous theoretical investigations \cite{Li-PRL-2009, Wang-PRL-2011, Daniele-PRB-2011, Zhao-PRL-2011, Maximilian-PRL-2012, Chen-PRL-2012, Zhou-PRB-2010, Umemoto-PRL-2010}, all of these candidate allotropes exhibit high hardnesses \cite{Lyakhov-PRB-2011} and bulk moduli, which are comparable with those of diamond. Fig. \ref{DOS} shows the calculated total and partial electronic densities of states in both systems. It can be clearly seen that 2\emph{p} states exhibit a larger overlap with 2\emph{s} states in diamond, which makes diamond the most stable allotrope among \emph{sp}$^3$ forms of carbon. The magnitude of overlap determines the order of stability: \emph{M}-carbon $>$ \emph{X}-carbon $>$ bct-C$_4$ $>$ \emph{Y}-carbon. The DFT band gaps of \emph{M}-carbon, \emph{X}-carbon, bct-C$_4$, and \emph{Y}-carbon are 3.6, 3.8, 2.7 and 2.9 eV, and we should bear in mind that DFT always underestimate the band gaps - so the real gaps are larger, and all of these $sp^3$-allotropes should be transparent colorless insulators.

\section{Conclusions}
In summary, we performed a systematic search for metastable allotropes of carbon that can be synthesized by cold compression of graphite by using the recently developed evolutionary metadynamics technique \cite{QZhu-metadynamics}. Starting from 2H- or 3R-graphite, at 20 GPa we easily found diamond as the ground state and observed a large variety of low-energy metastable \emph{sp}$^3$ carbon allotropes accessible from graphite. Apart from diamond, lonsdaleite and their polytypes, we summarize four other families of low-enthalpy carbon allotropes which can be obtained by cold compression of graphite, (i)5+7 topology (\emph{M/W}-carbon); (ii)5+7 topology (\emph{X}-carbon); (iii)4+8 topology (bct-C$_4$); (iv)4+8 topology (\emph{Y}-carbon). All of these structures are consistent with experimental data on `superhard graphite', and are predicted to have excellent mechanical properties, but transition path sampling calculations \cite{SEB-2012} unequivocally show \emph{M}-carbon to be the likeliest product of cold compression of graphite. Yet, the allotropes predicted here could be synthesized with a different experimental protocol and starting from different materials (graphite-3R, turbostratic graphites, fullerenes, or nanotubes, etc). We find it particularly encouraging that all the previously proposed structures and two topologically interesting and previously not described ones (X- and Y-carbon) were found in a systematic way, using just one calculation per starting material (graphite-2H or 3R). Our work shows that evolutionary metadynamics is a powerful method for efficiently finding not only stable but also low-energy metastable structures.

\begin{acknowledgments}
Calculations were performed at the supercomputer of Center for Functional Nanomaterials, Brookhaven National Laboratory, which is supported by the U.S. Department of Energy, Office of Basic Energy Sciences, under Contract No. DE-AC02-98CH10086, and at the Joint Supercomputer Center of the Russian Academy of Sciences and on the Skif supercomputer (Moscow State University). This work is funded by DARPA (No. N66001-10-1-4037), National Science Foundation (No. EAR-1114313). The research fund of the State Key Laboratory of Solidification Processing of NWPU, China (No. 65-TP-2011), and the Nature Science Foundation of China (No. 50802076 and 11174152) are also acknowledged. Evolutionary metadynamics has been implemented into the USPEX code, which is available at http://han.ess.sunysb.edu/~USPEX.
\end{acknowledgments}
\bibliography{1.bib}

\end{document}